
%
%
%
%
%
%
\font\bigrm=cmb10 scaled\magstep1
\font\srm=cmr9
\font\csc=cmcsc10

\def\foot{\baselineskip=12pt\srm\footnote}


\magnification=\magstep1
\vsize 8.9truein    \hsize 6.4truein 
\topskip 10pt       \leftskip 0pt  \rightskip 0pt
\baselineskip=19pt plus 2pt minus 1pt
\parskip 0pt plus 1pt    \parindent 20pt


\def\ha{{1 \over 2}}
\def\->{\rightarrow}     \def\<-{\leftarrow}
\def\<{\langle\,}          \def\>{\,\rangle}
\def\[{\left [\,}          \def\]{\,\right ]}
\def\({\left (\,}          \def\){\,\right )}
        
\def\|{\vert}

\def\der#1{{\partial \over \partial #1}}

\def\no{\noindent}
\def\hb{\hfill\break}
\def\qed{\ \ \ \ \hbox{\hfill\vbox{\hrule width8pt
\noindent\vrule height8pt\hskip 7.5pt\vrule height8pt\hrule width8pt}}}


\def\a{\alpha}

\def\d{\delta}

\def\la{\lambda}    \def\La{\Lambda}
      
\def\p{\phi}              
\def\ps{\psi}

\def\t{\theta}	        


\def\bZ{{\bf  Z}} \def\bC{{\bf  C}} 
   
\def\cO{{\cal O}} \def\cW{{\cal W}} 


\def\W{\cW_{1+\infty}}
\def\:{\,{ }^\circ_\circ\,}
\def\pb{\bar\psi}  \def\der{\partial}


\def\Sk{\hskip20pt}
\def\H{\hrulefill} \def\V{\vrule height20pt} \def\VV{\vrule height20pt}

\def\Fi{${}^{f_1}$} \def\Fj{${}^{f_2}$} \def\Fk{${}^{f_p}$}

\def\Gi{$\,\,\,\,{}^{g_1}$}
\def\Gj{$\,\,\,\,{}^{g_2}$}
\def\Gk{$\,\,\,\,{}^{g_q}$}

\def\Ni{${}^{n_1}$} \def\Nj{${}^{n_2}$} \def\Nk{${}^{n_t}$}

\def\Mi{$\,\,{}^{m_1}$} \def\Mj{$\,\,{}^{m_2}$} \def\Mk{$\,\,{}^{m_t}$}

\def\DD{$\,\,{}^{\ddots}$}
\def\CD{$\,\,\,{}^{\cdots}$}
\def\VD{${}^{\vdots}$}


\def\rBK{9}
\def\RfBK{I. Bakas and E. Kiritsis,
Nucl. Phys. {\bf B343} (1990) 185-204;
Mod. Phys. Lett. {\bf A5} (1990) 2039-2050.}

\def\rBPRSS{8}
\def\RfBPRSS{E. Bergshoeff, C. Pope, L. Romans, E. Sezgin and X. Shen,
Phys. Lett. {\bf B245} (1990) 447-452.}

\def\rBS{1}
\def\RfBS{See references in P. Bouwknegt and K. Schoutens,
Phys. Rept. {\bf 223} (1993) 183-276.}

\def\rCTZ{7}
\def\RfCTZ{A. Cappelli, C. Trugenberger and G. Zemba,
Nucl. Phys. {\bf B396} (1993) 465-490;
{\it Classification of quantum Hall universality classes
by $\W$ symmetry,}
Preprint MPI-PH-93-75, November 1993, (hepth/9310181).}

\def\rDJKM{14}
\def\RfDJKM{E. Date, M. Jimbo, M. Kashiwara and T. Miwa,
Proc. of RIMS Symposium on
Non-Linear Integrable Systems - Classical Theory and Quantum Theory,
(ed. M. Jimbo and T. Miwa), 39-120, World Sci., 1983.}

\def\rFKN{2}
\def\RfFKN{M. Fukuma, H. Kawai and R. Nakayama,
Commun. Math. Phys. (1991) 371-403.}

\def\rG{5}
\def\RfG{J. Goeree,
Nucl. Phys. {\bf B358} (1991) 737-757.}

\def\rIKS{6}
\def\RfIKS{S. Iso, D. Karabali and B. Sakita,
Phys. Lett. {\bf B296} (1992) 143-150.}

\def\rIM{3}
\def\RfIM{H. Itoyama and Y. Matsuo,
Phys. Lett. {\bf B262} (1991) 233-239.}

\def\rKR{12}
\def\RfKR{V. Kac and A. Radul,
{\it Quasi-finite highest weight modules over the Lie algebra of
differential operators on the circle,}
MIT Mathematics preprint, July 1993, (hepth/9308153).}

\def\rKS{4}
\def\RfKS{V. Kac and A. Schwarz,
Phys. Lett. {\bf B257} (1991) 329-334;\hb\no
A. Schwarz,
Mod. Phys. Lett. {\bf A6} (1991) 2713-2726.}

\def\rM{15}
\def\RfM{Y. Matsuo,
{\it Free fields and quasi-finite representation of $\W$ algebra,}
Preprint UT-661, December 1993, (hepth/9312192).}

\def\rO{11}
\def\RfO{S. Odake,
Int. J. of Mod. Phys. {\bf A7} (1992) 6339-6355.}

\def\rOS{10}
\def\RfOS{S. Odake and T. Sano,
Phys. Lett. {\bf B258} (1991) 369-374.}

\def\rPRS{13}
\def\RfPRS{C. Pope, L. Romans and X. Shen,
Nucl. Phys. {\bf B339} (1990) 191-221.}

\def\Fha{e-mail address : awata@ps1.yukawa.kyoto-u.ac.jp}
\def\Fmf{e-mail address : fukuma@ps1.yukawa.kyoto-u.ac.jp}
\def\Fso{e-mail address : odake@jpnyitp.yukawa.kyoto-u.ac.jp}
\def\Fyq{e-mail address : quano@kurims.kyoto-u.ac.jp}
{\baselineskip=12pt
\rightline{\vbox{\hbox{YITP/K-1049}
                 \hbox{SULDP-1993-1}
                 \hbox{RIMS-959 }
                 \hbox{December 1993}
}}
}

\vskip.25in
\centerline{\bigrm  Eigensystem and Full Character Formula }
\centerline{\bigrm  of the $\W$ Algebra with c=1 }
\vskip.5in\centerline{
{\csc Hidetoshi AWATA}{\foot{$^1$}\Fha} ,
{\csc Masafumi FUKUMA}{\foot{$^2$}\Fmf} ,
{\csc Satoru    ODAKE}{\foot{$^3$}\Fso}}
\centerline{\ and \
{\csc Yas-Hiro QUANO}{\foot{$^4$}\Fyq}
}

{\baselineskip=14pt
\it\vskip.25in
\centerline{$^{1,2}$ Yukawa Institute for Theoretical Physics}
\centerline{Kyoto University, Kyoto 606, Japan}
\vskip.1in
\centerline{$^3$ Department of Physics, Faculty of Liberal Arts}
\centerline{Shinshu University, Matsumoto 390, Japan}
\vskip.1in
\centerline{$^4$ Research Institute for Mathematical Sciences}
\centerline{Kyoto University, Kyoto 606, Japan}
}
\vskip.4in\centerline{\bf Abstract}\vskip.15in

By using the free field realizations,
we analyze the representation theory of the $\W$ algebra with $c=1$.
The eigenvectors for the Cartan subalgebra of $\W$ are
parametrized by the Young diagrams,
and explicitly written down by $\W$ generators.
Moreover, their eigenvalues and full character formula are also obtained.

\vskip.3in
hep-th/9312208

\vfill\eject

\vskip3mm
\no{\bf 1. Introduction }
\vskip2mm

The $\W$ algebra [{\rBS}] appears
in many two-dimensional physical systems,
for example, quantum gravity [{\rFKN}, {\rIM}, {\rKS}, {\rG}]
and quantum Hall effect [{\rIKS}, {\rCTZ}].
However, we are far from applying the $\W$ algebra to these physical systems,
since the representation theory of the $\W$ algebra
has not been understood enough.
On the other hand, we know that the $\W$ algebra
is realized by free fields [{\rBPRSS}],
as is the case for the ${\cW_{\infty}}$ algebra [{\rBK}]
and its generalizations [{\rOS}, {\rO}].
The aim of the present letter is to demonstrate that
the representation theory of the $\W$ algebra
is easily investigated by using this free field realization.
Although we here exclusively consider 
the case when the central charge of the $\W$ algebra is unity,
the generalization of our analysis to other central charges
is straightforward, and will be reported elsewhere
with the relation of our work with the one by Kac and Radul [{\rKR}].

This letter is arranged as follows.
In sect.\ 2, we introduce free fermions and free bosons
which play the fundamental role in the analysis of the $\W$ algebra.
We then in sect.\ 3 study the representation theory of the $\W$ algebra
on the basis of the fermionization.
This section contains the main results of the letter.
In sect.\ 4, we discuss the bosonization of the $\W$ algebra.
Sect.\ 5 is devoted to conclusions. 

\vskip3mm
\no{\bf 2. Free fermions, bosons and their correspondence}
\vskip2mm
\def\Efv{2.1.1}
\no{\bf 2.1.}~
We first fix some notations.
The free fermion fields 
$$
\pb(z)=\sum_{r\in\bZ-\ha}\pb_r z^{-r-\ha},\qquad
\ps(z)=\sum_{r\in\bZ-\ha}\ps_r z^{-r-\ha},
$$
are defined with the anti-commutation relations
$\{\pb_r,\ps_s\}=\d_{r+s,0}$, 
$\{\ps_r,\ps_s\} = \{\pb_r,\pb_s\} = 0$,
and thus satisfy the following OPE relations:
$$\eqalign{
\pb(z)\ps(w)&={ 1\over z-w}+\:\pb(z)\ps(w)\:,\cr
\ps(z)\pb(w)&={ 1\over z-w}+\:\ps(z)\pb(w)\:,
}\qquad\eqalign{
\ps(z)\ps(w)&=             \:\ps(z)\ps(w)\:,\cr
\pb(z)\pb(w)&=             \:\pb(z)\pb(w)\:.
}$$
Here $\{A,B\}=AB+BA$, and
$(1-x)^{-1}=\sum_{n\in\bZ\ge 0}x^n$.
We denote by $\:\,\cO\,\:$ the fermionic normal ordering defined as
$\:A_r B_s\: = A_r B_s \,\t(r<1/2) - B_s A_r \,\t(r\ge 1/2)$,
where $A_r$ and $B_r$ are $\ps_r$ or $\pb_r$,
and $\t(P)$ is a step function such that
$\t(P)=1$ if the statement $P$ is true, otherwise zero. 


The fermion Fock space $F$ is spanned by the vectors
$$
\pb_{-r_1}\cdots\pb_{-r_k} \ps_{-s_1}\cdots\ps_{-s_l}\|0\>,\qquad
r_i>r_{i+1}>0,\qquad s_i>s_{i+1}>0,
\eqno(\Efv)
$$
where $\|0\>$ is the highest weight vector such that
$\pb_r\|0\> = \ps_r\|0\> = 0 $ for $r>0$.


\vskip3pt
\no{\bf 2.2.}~
The free boson field 
$$
\p(z) = q + \a_0 \log z - \sum_{n\in\bZ_{\neq 0}}{\a_n \over n} z^{-n},
$$
is defined with the commutation relations
$\[\a_n,\a_m\] = n \d_{n+m,0}$, 
$\[\a_0,q\]=1$,
and thus satisfies the following OPE relation:
$$
\p(z)\p(w) = \log (z-w) + : \p(z)\p(w) :.
$$
Here $\[A,B\]=AB-BA$, and
$\log (1-x)=-\sum_{n\in\bZ> 0}x^n/n$.
We denote by $:\cO:$ the bosonic normal ordering defined as
$:\a_n \cO: \,= \a_n :\cO: \,\t(n<0)\, + :\cO: \a_n \,\t(n\geq 0)$ and
$:q\, \cO: =q :\cO: $,
where $\cO\in\bC[\a_n,q]$.


The boson Fock space $B(\La)$ is spanned by the vectors
$\a_{-m_1}\cdots\a_{-m_k}\|\La\>$,
with $m_i\ge m_{i+1}>0$,
where $\|\La\>$ with $\La\in\bC$ is the highest weight vector such that
$\a_n\|\La\> = 0 $ for $n>0$ and $\a_0\|\La\> = \La\|\La\>$.
Note that $\|\La\> = e^{\La q}\|0\>$.

\def\EN{2.3.1}
\vskip3pt
\no{\bf 2.3.}~
As is well known, there exists the fermion-boson correspondence.
The free fermion fields $\pb(z)$ and $\ps(z)$ are realized by
the free boson field $\p(z)$ as
$\pb(z)=: e^{ \p(z)}:$ and
$\ps(z)=: e^{-\p(z)}:$.
On the other hand,
the $U(1)$ current $\der \p(z)$ is realized by
the free fermion fields as the fermion number current
$\der \p(z)=\:\pb(z) \ps(z)\:$,
and the zero-mode operator $q$ plays the role of
the fermion number sift operator.
The highest weight vector $\|N\>$ for $N\in\bZ$ of boson Fock space $B(N)$
is then expressed as
$$
\|N\>=\left\{\eqalign{
&\pb_{-N+\ha}\cdots\pb_{-{3\over 2}}\pb_{-\ha}\|\,0\>,\cr
&                                             \|\,0\>,\cr
&\ps_{ N+\ha}\cdots\ps_{-{3\over 2}}\ps_{-\ha}\|\,0\>,
}\right.\qquad\eqalign{
&N>0,\cr &N=0,\cr &N<0.
}\eqno(\EN)
$$
Therefore, we have the relation 
$F = \oplus_{N\in\bZ}B(N)$.

\vskip5mm
\no{\bf 3. The $\W$ algebra}
\vskip2mm
\def\Fgl{
The $\W$ algebra spanned by the generators $W^k_n$ in eq.\ ({\Ew})
is a subalgebra of $gl(\infty)$. In particular,
the $W^1(z)$ is the time evolution generator of KP hierarchy [{\rDJKM}].
}
\def\Ew{3.1.1}
\def\Eww{3.1.2}

\no{\bf 3.1.}~
The $\W$ algebra with central charge $c=1$ [{\rPRS}] is
nothing but the algebra of ``local'' bilinears of fermions;
the generating currents $W^k(z)$ $(k\in\bZ_{>0})$
of the $\W$ algebra are defined as follows:
$$
W^k(z)=\sum_{p,q,t=0\atop p+q+t=k-1}^{k-1}
c^k_{p,q,t}\:\der^p\pb(z) \der^{q}\ps(z)\: z^{-t},
$$
with arbitrary constants $c^k_{p,q,t}\in\bC$.
Or equivalently,
$$
W^k(z)= \sum_{n\in\bZ} W^k_n z^{-n-k},\qquad
W^k_n = \sum_{r,s\in\bZ-\ha\atop r+s=n} c^k_{r,s} \:\pb_r\ps_s\:,
\eqno(\Ew)
$$
with{\foot{${}^\dagger$}\Fgl}
$$
c^k_{r,s}=\sum_{p,q,t=0\atop p+q+t=k-1}^{k-1}
c^k_{p,q,t}\[-r-\ha\]_p\[-s-\ha\]_q,
$$
where $\[n\]_m\equiv\prod_{j=0}^{m-1}(n-j)$.
If we set $c^1_{0,0,0}=1$, then $c^1_{r,s}=1$,
and thus $W^1(z)$ is just the fermion number current
which realized by the free boson as $\der\p(z)$.
Since
$$
\[\,W^k_n,\pb_{r}\,\] = c^k_{r+n,-r} \pb_{r+n},\qquad
\[\,W^k_n,\ps_{r}\,\] =-c^k_{-r,r+n} \ps_{r+n},
$$
the generators $W^k_n$ satisfy the $\W$ algebra
$$
\[W^k_n,W^s_m\] =
\sum_{\ell\ge 2} g^{k,s,\ell}_{n,m} W^{k+s-\ell}_{n+m}+C^{k,s}_{n}\d_{n+m,0},
\eqno(\Eww)
$$
with some constants $g^{k,s,\ell}_{n,m}$, $C^{k,s}_{n}\in\bC$.


For example, the standard basis for the $\W$ generators [{\rPRS}] is
$$
c^k_{p,q,t}=(-1)^q \({p+q\atop q}\)^2 \({2(k-1)\atop k-1}\)^{-1}\d_{t,0},
$$
with $\({n\atop m}\)=[n]_m/m!$.
In this basis,
the anomaly terms in eq.\ ({\Eww}) 
are diagonarized, that is to say $C^{k,s}_n\propto\d_{k,s}$.
Moreover, they preserve the hermiticity;
namely, if we set $\ps_r^{\dagger}=\pb_{-r}$,
then  $W^{k\dagger}_n=W^k_{-n}$.
Since the $\W$ algebra is a Lie algebra,
any basis obtained by the invertible linear transformation
from this standerd basis as
$\tilde W^k_n = \sum_{s\ge 1} T^k_s\, W^s_n$ with $T^k_s\in\bC$
generate the same $\W$ algebra.

\vskip3pt
\no{\bf 3.2.}~
It is easy to see that the vector $\|N\>$ in eq.\ ({\EN}) is
the highest weight vector of the $\W$ algebra
which satisfies $W^k_n\|N\>=0$ for $n>0$ and $W^1_0\|N\>=N\|N\>$.
Let then $M(N)$ and $L(N)$ be, respectively,
the Verma module and the irreducible module over the $\W$ algebra
with respect to the highest weight vector $\|N\>$.
Since any generators of the $\W$ algebra and
the highest weight vector $\|N\>$ are realized by fermions,
we have the relation $F\supset\sum_{N\in\bZ}M(N)$.
Note here that $M(N)\cap M(N')=\emptyset$ for $N\neq N'$,
since $\[W^1_0, W^k_n\]=0$.
Thus, we obtain the relation $F\supset\oplus_{N\in\bZ}M(N)$.
On the other hand, since the oscillator modes $\a_n$ of
the free boson field belong to the $\W$ algebra,
we also have the relation
$\oplus_{N\in\bZ}B(N)\subset\oplus_{N\in\bZ}M(N)$.
Thus, we conclude that $F = \mathop\oplus_{N\in\bZ}M(N)$.
Futhermore,
since the Kac determinant for 
the fermion Fock space $F$ does not vanish,
the Verma module $M(N)$ is irreducible, {\it i.e.} $M(N)=L(N)$.
We thus have proved the following theorem [{\rO}].

\vskip3pt
\no{\bf\csc Theorem 3.2}.~{\it
The fermion Fock space $F$ is the direct sum of
the irreducible modules $L(N)$ $(N\in\bZ)$ over $\W$:
$$
F=\mathop\oplus_{N\in\bZ}L(N).
$$
}

\def\Ef{3.3.1}
\def\Ech{3.3.2}
\vskip3pt
\no{\bf 3.3.}~
The Cartan subalgebra of $\W$ is spanned by $W^k_0$,
for which the following equation holds:
$$
\[\,W^k_0,\pb_{r}\,\] = \bar a^k_{r} \pb_{r},\qquad
\[\,W^k_0,\ps_{r}\,\] =      a^k_{r} \ps_{r},
\eqno(\Ef)
$$
with $\bar a^k_{r} = c^k_{r,-r}$ and $a^k_{r} =-c^k_{-r,r}$.
Therefore, any vector in the form of eq.\ ({\Efv})
is an eigenvector of the Cartan subalgebra.
%
%
Hence, the full character ${\rm ch} L(N)$
of the $\W$ algebra,
$$
{\rm ch} L(N) = {\rm tr}_{L(N)} \prod_{k\ge 1} x_k^{W^k_0},
$$
is now easily calculated
by taking a trace over the fermion Fock space
$F = \mathop\oplus_{N\in\bZ}L(N)$ as follows:

\vskip3pt
\no{\bf\csc Theorem 3.3}.~{\it
The generating function of the full characters ${\rm ch}L(N)$ is
$$
\sum_{N\in\bZ} z^N{\rm ch}L(N)
=\prod_{r-\ha\in\bZ_{\ge 0}}
\( 1 + z      \prod_{k\ge 1} x_k^{\bar a^k_{-r}} \)
\( 1 + z^{-1} \prod_{k\ge 1} x_k^{     a^k_{-r}} \).
\eqno(\Ech)
$$
}
The character formula with $x_k=1$ for $k\ge 3$
was obtained in Ref.\ [{\rO}].

The right hand side of eq.\ ({\Ech}) can be rewritten as
$$
\exp\left\{\sum_{n\ge 1}{(-1)^{n-1}\over n}\bar f(x_k^n)z^{ n}\right\}\,
\exp\left\{\sum_{n\ge 1}{(-1)^{n-1}\over n}     f(x_k^n)z^{-n}\right\},
$$
where
$$
\bar f(x_k)=\sum_{r-\ha\in\bZ_{\ge 0}}\prod_{k\ge 1}x_k^{\bar a^k_{-r}},\qquad
     f(x_k)=\sum_{r-\ha\in\bZ_{\ge 0}}\prod_{k\ge 1}x_k^{     a^k_{-r}}.
$$
Thus, by introducing the elementary Schur polynomials as
$$
\eqalign{\sum_{n\in\bZ}\bar P_n z^n &=
\exp\left\{\sum_{n\ge 1}{(-1)^{n-1}\over n}\bar f(x_k^n)z^{ n}\right\},\cr
\sum_{n\in\bZ}     P_n z^n &=
\exp\left\{\sum_{n\ge 1}{(-1)^{n-1}\over n}f(x_k^n)z^{-n}\right\},
}$$
the full character ${\rm ch}L(N)$ is now expressed as
$$
{\rm ch} L(N)=\sum_{n,m\in\bZ\atop n+m=N}\bar P_n P_m.
$$

\def\Ff{We have inserted a phase factor for later convenience.}
\def\Eeva{3.4.1}
\def\Eev{3.4.2}
\vskip3pt
\no{\bf 3.4.}~
We next discuss eigenvectors and eigenvalues for
the Cartan subalgebra in $L(N)$.
Due to eq.\ ({\Ef}), we know that
the eigenvectors in $L(N)$ have the following form:
$$
\pb_{-r_1}\cdots\pb_{-r_t} \ps_{-s_1}\cdots\ps_{-s_t}\|N\>,
\qquad r_i>r_{i+1}>N,\quad s_i>s_{i+1}>-N.
$$
Here the fermion number of the above state is $N$,
since $\pb$ and $\ps$ appear the same times.
The eigenvalues are easily calculated by using eq.\ ({\Ef}),
and we have the following theorem.

\vskip3pt
\no{\bf\csc Theorem 3.4}.~{\it 
Eigenvectors of the Cartan subalgebra of $\W$ in $L(N)$
are parametrized by ordered sets
$Y=\(n_1>\cdots>n_t\,\|\,m_1>\cdots>m_t\)$ with $n_i$,
$m_i\in\bZ_{\ge 0}${\rm :}{\foot{${}^\dagger$}\Ff}
$$
\|N;Y\>=
 \pb_{-N-n_1-\ha}\cdots\pb_{-N-n_t-\ha}
 \ps_{ N-m_1-\ha}\cdots\ps_{ N-m_t-\ha}\|N\>(-1)^{\sum_{i=1}^t (m_i+i-1)},
$$
and the eigenvalue of $\|N;Y\>$ is
$$
W^k_0 \|N;Y\>=\( w^k_N + Y^k_N(Y)\) \|N;Y\>,
\eqno(\Eeva)
$$
with
$$\eqalign{
w^k_N&=
    \sum_{r=\ha}^{ N-\ha} \bar a^k_{-r}\,\t(N\ge 0)
  + \sum_{s=\ha}^{-N-\ha}      a^k_{-s}\,\t(N <  0),\cr
Y^k_N(Y)&=\sum_{i=1}^t \(\bar a^k_{-N-n_i-\ha} + a^k_{N-m_i-\ha}\),
}$$
where $w^k_N$ is the weight of the highest weight vector $\|N\>$.
}


Besides the above parametrization $Y$,
we have other expressions for the eigenvectors, for example,
$$\eqalign{
\pb_{-N-f_1+\ha}\cdots\pb_{-N-f_p+p-\ha}&\|N-p\>,\qquad
f_i\ge f_{i+1}\in\bZ_{>0},\cr
\ps_{ N-g_1+\ha}\cdots\ps_{ N-g_q+q-\ha}&\|N+q\>,\qquad
g_i\ge g_{i+1}\in\bZ_{>0}.
}\eqno(\Eev)
$$
Such many different expressions for the same eigenvector are
understood as the different parametrizations for the same Young diagram.
%
%
The first expression for the eigenvectors in the theorem 3.4
corresponds to the following parametrization for the Young diagram
$Y=\(n_1>\cdots>n_t\,\|\,m_1>\cdots>m_t\)$ with $n_i$, $m_i\in\bZ_{\ge 0}$:
$$
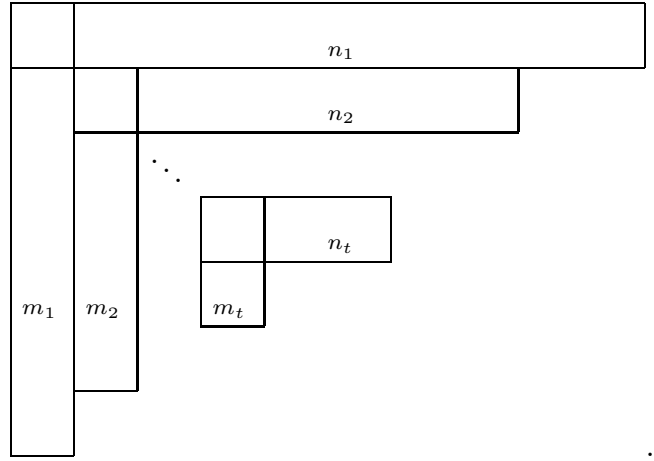
\vbox{
\offinterlineskip
\settabs
\+\Sk  &\Sk  &\Sk  &\Sk  &\Sk  &\Sk  &\Sk  &\Sk  &\Sk  &\Sk  &\Sk\cr
\+\H   &\H   &\H   &\H   &\H   &\H   &\H   &\H   &\H   &\H   &   \cr
\+\V   &\V   &     &     &     &\Ni  &     &     &     &     &\V \cr
\+\H   &\H   &\H   &\H   &\H   &\H   &\H   &\H   &\H   &\H   &   \cr
\+\V   &\V   &\V   &     &     &\Nj  &     &     &\V   &     &   \cr
\+     &\H   &\H   &\H   &\H   &\H   &\H   &\H   &     &     &   \cr
\+\VV  &\VV  &\VV\DD&    &     &     &     &     &     &     &   \cr
\+     &     &     &\H   &\H   &\H   &     &     &     &     &   \cr
\+\VV  &\VV  &\VV  &\VV  &\VV  &\Nk  &\VV  &     &     &     &   \cr
\+     &     &     &\H   &\H   &\H   &     &     &     &     &   \cr
\+\VV\Mi&\VV\Mj&\VV&\VV\Mk&\VV &     &     &     &     &     &   \cr
\+     &     &     &\H   &     &     &     &     &     &     &   \cr
\+\V   &\V   &\V   &     &     &     &     &     &     &     &   \cr
\+     &\H   &     &     &     &     &     &     &     &     &   \cr
\+\V   &\V   &     &     &     &     &     &     &     &     &   \cr
\+\H   &     &     &     &     &     &     &     &     &     &   \cr
}.$$
%
The other two expressions for the eigenvectors in eq.\ ({\Eev})
correspond to the following parametrizations for the Young diagram
$Y=\(f_1\ge\cdots\ge f_p\)$ or $Y=\(g_1\ge\cdots\ge g_q\)$
with $f_i$, $g_i\in\bZ_{>0}$, respectively:
$$
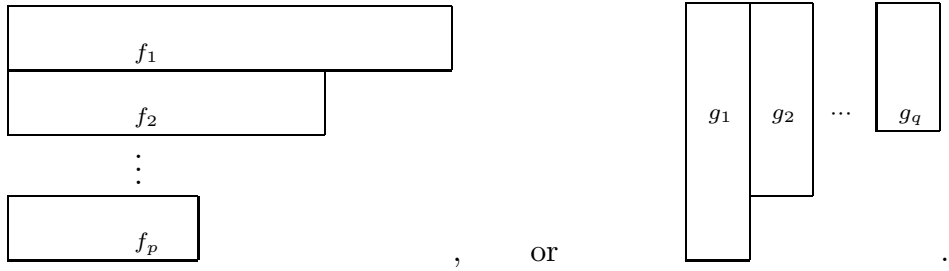
\vbox{
\offinterlineskip
\settabs
\+\Sk  &\Sk  &\Sk  &\Sk  &\Sk  &\Sk  &\Sk  &\Sk  \cr
\+\H   &\H   &\H   &\H   &\H   &\H   &\H   &     \cr
\+\V   &     &\Fi  &     &     &     &     &\V   \cr
\+\H   &\H   &\H   &\H   &\H   &\H   &\H   &     \cr
\+\V   &     &\Fj  &     &     &\V   &     &     \cr
\+\H   &\H   &\H   &\H   &\H   &     &     &     \cr
\+     &     &\VD  &     &     &     &     &     \cr
\+\H   &\H   &\H   &     &     &     &     &     \cr
\+\V   &     &\Fk  &\V   &     &     &     &     \cr
\+\H   &\H   &\H   &     &     &     &     &     \cr
},\qquad{\rm or}\qquad\qquad\vbox{
\offinterlineskip
\settabs
\+\Sk   &\Sk   &\Sk   &\Sk   &\Sk  \cr
\+\H    &\H    &      &\H    &     \cr
\+\VV   &\VV   &\VV   &\VV   &\VV  \cr
\+\VV\Gi&\VV\Gj&\VV\CD&\VV\Gk&\VV  \cr
\+      &      &      &\H    &     \cr
\+\VV   &\V    &\V    &      &     \cr
\+      &\H    &      &      &     \cr
\+\V    &\V    &      &      &     \cr
\+\H    &      &      &      &     \cr
}.$$

\vskip3pt
\no{\bf 3.5.}~
We now calculate the eigenvalues explicitly.
To do so, we introduce a new basis of $\W$ for
which the eigenvalues have a simple form;
$$
c^{2k-1}_{p,q,t}=(-1)^{k-1}\d_p^{k-1}\d_q^{k-1} \d_t^{0},\qquad
c^{2k  }_{p,q,t}={(-1)^{k-1}\over 2}
\(\d_p^{k}\d_q^{k-1} - \d_p^{k-1}\d_q^{k}\) \d_t^{0},
$$
or equivalently,
$$\eqalign{
W^{2k-1}(z)&=(-1)^{k-1}\:\der^{k-1}\pb(z) \der^{k-1}\ps(z)\:,\cr
W^{2k  }(z)&={(-1)^{k-1}\over 2}
\:\(\der^{k}\pb(z)\der^{k-1}\ps(z)-\der^{k-1}\pb(z)\der^{k}\ps(z)\)\:.
}$$
For this basis, one can easily obtain
$$
\bar a^{2k-1}_r =   \prod_{s={3\over 2}-k}^{k-{3\over 2}}(r+s),\qquad
\bar a^{2k  }_r =r\,\prod_{s={3\over 2}-k}^{k-{3\over 2}}(r+s),\qquad
     a^k_r =(-1)^k \bar a^k_r.
$$
Thus, the eigenvalue of
the eigenvector $\|N;Y\>$ is evaluated as eq.\ ({\Eeva}) with
$$
w^{2k-1}_N = {1\over 2k-1} \prod_{\ell=1-k}^{k-1} (N+\ell), \qquad
w^{2k  }_N = {N\over 2k}   \prod_{\ell=1-k}^{k-1} (N+\ell),
$$
and
$$\eqalign{
&Y^{2k-1}_N(Y) =(2k-2)\sum_{(i,j)\in Y}\prod_{\ell=2-k}^{k-2}(N+j-i+\ell),\cr
&Y^{2k  }_N(Y) =(2k-1)\sum_{(i,j)\in Y}
(N+j-i) \prod_{\ell=2-k}^{k-2} (N+j-i+\ell),\qquad
Y^{2  }_N(Y) =\sum_{(i,j)\in Y}1,
}$$
where $(i,j)\in Y$ means that the Young diagram $Y$ has a box
in the place of the $i$-th row and $j$-th column.
Here we have used the identities
$$\eqalign{
\sum_{m=a}^b   \prod_{n=-c}^c (m+n)
&= {1\over 2c+2 }
\left\{ \prod_{n=-c}^{c+1} (b+n) - \prod_{n=-c-1}^{c} (a+n) \right\},\cr
\sum_{m=a}^b m \prod_{n=-c}^c (m+n)
&= {1\over 2c+3 }
\left\{ (b+\ha) \prod_{n=-c}^{c+1} (b+n)
      - (a-\ha) \prod_{n=-c-1}^{c} (a+n) \right\}.
}$$

\vskip5mm
\no{\bf 4. Bosonization of the $\W$ algebra }
\vskip2mm

\no{\bf 4.1.}~
We now study the representation theory of
the $\W$ algebra in terms of the free boson field.
Since
$$\eqalign{
\:\der^p\pb(z)\der^q\ps(z)\: &= \sum_{m,n\in\bZ_{\ge 0}\atop m+n=p+q}
b^{p,q}_{m,n} \der^m P^{(n+1)}(z),\cr
P^{(n)}(z) = \,:e^{-\p(z)}\der^n e^{\p(z)}:\,
&= \,:\( \der + \der\p(z) \)^n \cdot 1:,\qquad
b^{p,q}_{m,n}={(-1)^{q-m}\over n+1}\({q\atop m}\),
}$$
the generators of the $\W$ algebra are realized by
the free boson field as follows [{\rFKN}]:
$$
W^{k}(z)= \sum_{m,n,t=0\atop m+n+t=k-1}^{k-1}
\tilde c^k_{m,n,t} \der^m P^{(n+1)}(z) z^{-t},
$$
with
$\tilde c^k_{m,n,t}=
\sum_{p,q=0}^{k-1}c^k_{p,q,t}b^{p,q}_{m,n}\d_{p+q+t,k-1}$.
Obviously, $B(N) = L(N)$.

\vskip3pt
\no{\bf 4.2.}~
Let $\bar S_n$ and $S_n$ with $n\in\bZ$ be
the elementary Schur polynomials defined by
$$
\sum_{n\in\bZ}\bar S_n z^n =
\exp\left\{ \sum_{n>0}{W^1_{-n}\over n}z^n \right\},
\qquad
\sum_{n\in\bZ} S_n z^n =
\exp\left\{-\sum_{n>0}{W^1_{-n}\over n}z^n \right\},
$$
and let $\chi_{m,n}$ with $m,n\in\bZ$ be the following operators:
$$
\chi_{m,n}
\equiv (-1)^{m+1}\sum_{\ell\ge 0} \bar S_{n-\ell  } S_{m+\ell+1}
=(-1)^{m  }\sum_{\ell\ge 0} \bar S_{n+\ell+1} S_{m-\ell  }.
$$
Then we have the following theorem.

\vskip3pt
\no{\bf\csc Theorem 4.2}.~{\it
The eigenvectors $\|N;Y\>$ associated with the Young diagram
$Y=$ \break
$\(n_1>\cdots>n_t\,\|\,m_1>\cdots>m_t\)$ with $n_i$, $m_i\in\bZ_{\ge0}$
is realized 
as
$$
\|N;Y\>
= \det\(\chi_{m_i,n_j} \)_{1\le i,j\le t}\|N\>.
$$
}

\no{\it Proof }.~
We consider the generating function of the eigenvectors,
$$
\pb(z_1)\cdots\pb(z_t) \ps(w_1)\cdots\ps(w_t)\|N\>
=\sum_{\{r_i,s_i\}}\pb_{r_1}\cdots\pb_{r_t} \ps_{s_1}\cdots\ps_{s_t}\|N\>
\prod_{i=1}^t z_i^{-r_i-\ha} w_i^{-s_i-\ha}.
$$
Note that the left hand side can be rewritten in terms of bosons as
$$
{\prod_{i<j} (z_i-z_j) (w_i-w_j) \over \prod_{i,j} (z_i-w_j)}
 :\prod_{i=1}^t e^{\p(z_i)} \prod_{j=1}^t e^{-\p(w_i)}: \|N\>.
$$
Thus, the theorem is obtained if we use the following identity:
$$
{}\qquad\qquad\qquad\quad
{\prod_{i<j} (z_i-z_j) (w_i-w_j) \over \prod_{i,j} (z_i-w_j)}
= (-1)^{\ha t(t-1)} \det\({1\over z_i-w_j}\)_{1\le i,j\le t}.
\qquad\qed
$$
Note that the polynomial $\det\(\chi_{m_i,n_j} \)_{1\le i,j\le t}$
is exactly the character polynomial appearing in Ref.\ [{\rDJKM}]
as the $\tau$ function of the KP hierarchy.

\vskip3pt
\no{\bf 4.3.}~
If we consider the other two parametrizations for the eigenvectors
in eq.\ ({\Eev}),
then we obtain the following proposition.

\vskip3pt
\no{\bf\csc Proposition 4.3}.~{\it
The eigenvectors $\|N;Y\>$ associated with the Young diagram
$Y=\(f_1\ge\cdots\ge f_p\)$ or $Y=\(g_1\ge\cdots\ge g_q\)$
with $f_i$, $g_i\in\bZ_{>0}$
is realized by the Schur polynomials of bosons as follows:
$$\eqalign{
\|N;Y\>
&= \det\( \bar S_{f_i+j-i}\)_{1\le i,j\le p}\|N\>.\cr
&= \det\( S_{g_i+j-i}\)_{1\le i,j\le q}\|N\>(-1)^{\sum_i g_i}.\cr
}$$
}

\no{\it Proof }.~
We now bosonize the generating function of the eigenvectors as
$$
\pb(z_1)\cdots \pb(z_p) \|N-p\>
 = \prod_{i<j} (z_i-z_j) :\prod_{i=1}^p e^{\p(z_i)}: \|N-p\>.
$$
By using the Vandermonde's determinant
$$
\prod_{i<j} (z_i-z_j)
= (-1)^{\ha p(p-1)} \det\(z_i^{j-1}\)_{1\le i,j\le p},
$$
we obtain the first part of the proposition.
The second part is proved similarly.
\qed



\bigbreak\no{\bf 5. Conclusion and Discussion}

On the basis of the free fermion realization,
we have identified the eigenvectors and eigenvalues
for the Cartan subalgebra of $\W$ with $c=1$,
which are parametrized by the Young diagrams.
Furthermore, we have obtained the full character formula
for the $\W$ algebra.

In addition, we wish to make several further comments.

First, in the case of the free boson realization,
not only the vector $\|N\>$ with $N\in\bZ$
but also $\|\La\>$ with $\La\in\bC$
is the highest weight vector of the $\W$ algebra.
Since $N$ is treated as an indeterminate variable in deriving
the formulas of characters ${\rm ch}L(N)$ and
eigenvectors $\|N;Y\>$,
we can replace $N\in\bZ$ by $\La\in\bC$ in the formulas.

Second, the $\W$ algebra can also be realized by
the $bc$ system with spins $\la$ and $1-\la$,
$$
b(z)=\sum_{r\in\bZ-\la  } b_r z^{-r-\la},\qquad
c(z)=\sum_{r\in\bZ+\la} c_r z^{-r+\la-1}.
$$
In fact, it is achieved simply by replacing
$\pb_r$ and $\ps_s$ with $b_{r+\ha-\la}$ and $c_{s-\ha+\la}$, respectively.
Eigenvectors and eigenvalues for the $bc$ system are
the same as those of the free fermion system.
If we redefine the Virasoro generator $W^2(z)$
by adding the derivative of $U(1)$ current $W^1(z)$,
then the central charge of the Virasoro generator varies.
However, the $\W$ algebra itself does not change
because this redefinition is nothing but
a linear transformation of the $\W$ generators $W^k_n$.

Third, even for the other quasi-finite case $c\neq 1$ [{\rKR}, {\rM}], %
we expect that the eigenvectors are also parametrized by the Young diagrams.

Finally, for other $W$ infinity algebras considered in Ref.\ [{\rO}],
we can easily write down
the full characters or the generating functions of them as the theorem 3.3.


\bigbreak\no{\bf Acknowledgments}

 The authors would like to thank Y. Matsuo
for valuable discussions.
H.A. and Y.-H.Q. are partly supported by Soryushi-syougakkai and
Grant-in-Aid for Scientific Research from the Ministry of Education,
Science and Culture (No. 04-2297), respectively.

\vskip 3mm
\no{\bf References}
\vskip2mm
\item{ }{ }
\vskip-0.5truecm
\baselineskip=14pt

\item{[\rBS]}{\RfBS}

\item{[\rFKN]}{\RfFKN}

\item{[\rIM]}{\RfIM}

\item{[\rKS]}{\RfKS}

\item{[\rG]}{\RfG}

\item{[\rIKS]}{\RfIKS}

\item{[\rCTZ]}{\RfCTZ}

\item{[\rBPRSS]}{\RfBPRSS}

\item{[\rBK]}{\RfBK}

\item{[\rOS]}{\RfOS}

\item{[\rO]}{\RfO}

\item{[\rKR]}{\RfKR}

\item{[\rPRS]}{\RfPRS}

\item{[\rDJKM]}{\RfDJKM}

\item{[\rM]}{\RfM}


\bye